\documentclass[12pt]{article}

\usepackage{amssymb,amsmath}

\DeclareMathOperator{\sech}{sech}
\DeclareMathOperator{\csch}{csch}

\begin{document}

\begin{titlepage}

\begin{flushright}
arXiv:2006.06793
\end{flushright}
\vskip 2.5cm

\begin{center}
{\Large \bf Separability of the Planar $1/\rho^{2}$ Potential\\
In Multiple Coordinate Systems}
\end{center}

\vspace{1ex}

\begin{center}
{\large Richard DeCosta and Brett Altschul\footnote{{\tt altschul@mailbox.sc.edu}}}

\vspace{5mm}
{\sl Department of Physics and Astronomy} \\
{\sl University of South Carolina} \\
{\sl Columbia, SC 29208} \\
\end{center}

\vspace{2.5ex}

\medskip

\centerline {\bf Abstract}

\bigskip

With a number of special Hamiltonians, solutions of the Schr\"{o}dinger equation may be found by separation
of variables in more than one coordinate system.
The class of potentials involved includes a number of important examples, including
the isotropic harmonic oscillator and the Coulomb potential. Multiply separable Hamiltonians exhibit a number of
interesting features, including ``accidental'' degeneracies in their bound state spectra and
often classical bound state orbits that always close. We examine another potential, for which the Schr\"{o}dinger equation
is separable in both cylindrical and parabolic coordinates: a $z$-independent $V\propto 1/\rho^{2}=1/(x^{2}+y^{2})$ in
three dimensions. All the persistent, bound classical orbits in this potential close, because all other orbits
with negative energies fall to the center at $\rho=0$. When separated in parabolic coordinates, the
Schr\"{o}dinger equation splits into three individual equations, two of which are equivalent to the radial equation
in a Coulomb potential---one equation with an attractive potential, the other with an equally strong repulsive potential.

\bigskip

\end{titlepage}

\newpage

\section{Introduction}

There are certain special Hamiltonians for which the Schr\"{o}dinger equation is separable in more than one
coordinate system. The spectra of these Hamiltonians exhibit what are known as ``accidental'' degeneracies, and
a number of the Hamiltonians are
extremely important. Well known examples of systems that are separable in more than one set of coordinates
include the free particle (separable in any coordinates for which the Robertson condition is
satisfied~\cite{ref-robertson,ref-morse}),
the three-dimensional isotropic harmonic oscillator (separable in spherical,
ellipsoidal, cylindrical, and rectangular coordinates),
the Coulomb potential (separable in spherical, prolate ellipsoidal~\cite{ref-coulson}, and parabolic coordinates),
and the constant magnetic field (separable in
rectangular or cylindrical coordinates, depending on the gauge, but in either case with a free choice of the location of the
origin). All these examples are also well known for the degeneracies present in their spectra. A slightly less well known
system is a particular anisotropic harmonic oscillator, with potential
\begin{equation}
\label{eq-anisoHO}
V(\vec{r}\,)=\frac{1}{2}M\omega_{0}^{2}(x^{2}+y^{2}+4z^{2})=\frac{1}{2}M\omega_{0}^{2}(r^{2}+3z^{2}).
\end{equation}
This is a special case of the kinds axially-symmetric harmonic oscillator potentials that are frequently useful in the
modeling of molecular vibrations in the presence of external fields~\cite{ref-petreska1,ref-petreska2}.
(Asymmetric oscillators may also be modeled by an extension of the harmonic oscillator formalism to an nonintegral
number of effective dimensions~\cite{ref-sandev}.)

That the energy spectrum with the specific potential (\ref{eq-anisoHO})
possesses accidental degeneracy is obvious; however, it is not as well known that,
just like the Coulomb problem, this potential problem is separable in parabolic coordinates.
Separability in parabolic coordinates is guaranteed if the Hamiltonian involves just a
standard nonrelativistic kinetic term and
a potential $V$ of the form
\begin{equation}
V(\vec{r}\,)=\frac{1}{r}\left[f(\eta)+g(\xi)\right],
\end{equation}
for some functions $f$ and $g$. The quantities $\eta$ and $\xi$ are two of the parameters of the parabolic coordinate
system $(\eta,\xi,\phi)$, where~\cite{ref-pauling}
\begin{eqnarray}
\eta & = & r+z \\
\xi & = & r-z.
\end{eqnarray}
(Surfaces of constant $\eta$ or $\xi$ are orthogonal paraboloids of revolution, each with its focus at the origin.)
For the anisotropic harmonic oscillator (\ref{eq-anisoHO}), the two functions are
$f(s)=g(s)=\frac{1}{4}M\omega_{0}^{2}s^{3}$.
The use of parabolic coordinates in the Coulomb problem is also especially convenient for dealing with the Stark
effect. While
the Hamiltonian remains separable in spherical coordinates when an external magnetic field is applied, it remains separable
in the parabolic coordinates with a external electric field present.

Classically, all the potentials we have mentioned with accidental degeneracy are well known for another feature.
All their bound orbits
close. In this paper, we shall look at another potential that, in a sense, shares this classical feature. Bertrand's
theorem is normally taken to hold that the only two central potentials for which all the
bound orbits close are the attractive Coulomb
potential and the isotropic harmonic oscillator potential. However, this is not quite accurate; there are other examples of
potentials with the stated property, but they are typically discounted because they do not have a full spectrum of
bound states.
For example, all the bounds orbits in a constant potential, which exerts no force,
close---precisely because there are no
bound states.
A system with a charged particle in a constant magnetic field also evades the strong restriction imposed by Bertrand's
Theorem~\cite{ref-bertrand},
since the force in this instance is not derived from a central potential. The orbits in the magnetic field are right
circular
helices. The velocity parallel to the magnetic field is a constant of the motion; only when the velocity in that direction
vanishes
are the orbits truly bound---in which case they are closed circles. The potential considered in this paper is in a similar
category to
the two examples just mentioned. We shall see that the
attractive $V\propto1/\rho^{2}$ potential possesses few bound orbits (for an appropriate
interpretation of the meaning of ``bound''), but those that it does possess are circular, with exactly zero energy
(as expected from the virial theorem).

Separability of the Schr\"{o}dinger equation
in a given set of coordinates means that the energy of a given state may be expressed as a function of
three quantum numbers, one corresponding to each coordinate. If separations in multiple coordinate systems are
possible, there
must be multiple formulas for the energy, based on different sets on quantum numbers. The eigenstates with fixed
quantum numbers are generally different in different coordinates; an eigenfunction in one coordinates must be a
linear combination
of eigenfunctions in the other coordinates. The only way this is possible for an energy eigenstate is if there are multiple
eigenstates with exactly the same energy. This shows why separability in multiple coordinate systems requires the
presence of
accidental degeneracy. Moreover, it is worth noting that the same argument can be applied even to the degeneracy of
system with a
generic central potential
$V(r)$, which is separable in spherical coordinates only. A system with angular momentum $\ell$ possesses a $(2\ell+1)$-fold
degeneracy, which is actually related to the fact that the spherical coordinates may be chosen with their polar
axis pointing in any direction. The Schr\"{o}dinger equation is thus separable in an infinite number of different
spherical coordinate systems.

However, the existence of accidental degeneracy does not absolutely require that a system be separable in multiple
coordinate systems. Any anisotropic three-dimensional harmonic oscillator for which the frequencies of the motions along
the three coordinate directions are rational multiples will have classical orbits that eventually close and degeneracies in
its quantum mechanical
spectrum. So there are systems for which the degeneracy is seemingly too ``sporadic'' to be indicative of any deeper
underlying symmetry principle at work.

The existence of alternative bases of quantum numbers is also related, of course, to the existence of additional
observables
that commute with the Hamiltonian. For the spherical harmonic oscillator and the Coulomb potential, these extra conserved
quantities are well known. The harmonic oscillator has separate, commuting Hamiltonians governing the motion along the
three orthogonal axes. For the Coulomb problem, there is the Runge-Lenz vector, which points out the direction of the major
axis of a bound state elliptical orbit. For the free particle, with its extensive degeneracy, the additional conserved
quantity is the momentum itself. Using the algebras generated with the inclusion of any of these conserved quantities,
it is possible to determine the bound state spectra of these problems using operator algebra alone.

In the Coulomb problem, part of the
accidental degeneracy associated with the additional operators that commute with the Hamiltonian
persists even in the relativistic Dirac theory, although
the separability in parabolic coordinates actually does not carry over. Another potential, albeit a potential
in only one dimension, that is also amenable to similar
operator methods is the $V=V_{0}\sech^{2}ax$ potential; the eigenstates for
potentials of different depths $V_{0}$ are related by operators~\cite{ref-gendenshtein},
and as with the other potentials solvable by operator 
methods, the operators involved may be interpreted as elements of a $(0+1)$-dimensional supersymmetry
algebra~\cite{ref-infield,ref-cooper2,ref-arai,ref-cooper1}. (An excellent introductory treatment of quantum-mechanical
supersymmetry,
especially as applied to the Coulomb problem, may be found in the lecture notes, Ref.~\cite{ref-rajagopal}.)
The harmonic oscillator, Coulomb, and $\sech^{2}ax$ potentials just mentioned are all among the shape-invariant
potentials, and the main object of study in this paper will be yet another such one.
There are a wide variety of mathematical tools that may be useful for solving and addressing questions about these
shape-invariant potentials~\cite{ref-kostelecky1,ref-cooper3}.

The harmonic oscillator and Coulomb systems are also known for the fact that certain classical and semiclassical
approximations yield exact results when applied
to these systems. The Bohr-Sommerfeld quantization rule derived from the Wentzel-Kramers-Brillouin (WKB) approximation
gives the
exact energies for a harmonic oscillator system in one dimension---and hence also for an isotropic or anisotropic harmonic
oscillator
in any number of dimensions. Harmonic oscillators also have coherent states, with the zero-point uncertainties in position
and momentum
added onto a classically orbiting wave function centroid. For the Coulomb potential, there is the fact that the full
nonperturbative scattering cross section is the same as the cross sections derived classically or from the first-order Born
approximation. Moreover, for the $V=V_{0}\sech^{2}ax$ potential also,
a certain approximation is exact; the potential is reflectionless, so the classical reflection coefficient is
always precisely correct.
The exact successes of these various approximations are highly appealing features of these special potential types,
although it is not clear whether we should expect anything similar when dealing with more esoteric multiply separable
potentials.

This paper will examine the $1/\rho^{2}=1/(x^{2}+y^{2})$ potential in three spatial dimensions. The attractive version
of the potential has been observed physically, in the interaction of a long charged wire with a polarizable
atoms and molecules~\cite{ref-denschlag,ref-nowak,ref-strebel}. For an atom with polarizability $\alpha$ in a spatially
slowly varying external electric field $\vec{E}_{0}$, the interaction energy is $-\frac{1}{2}\alpha\vec{E}_{0}^{2}$.
When the atom is exposed to the electrostatic field of a charged wire carrying linear charge density $\lambda$, the
$\vec{E}_{0}=\frac{2\lambda}{\rho}\hat{\rho}$ behavior of the field gives rise to an effective potential
\begin{equation}
V=-\frac{2\alpha\lambda^{2}}{\rho^{2}}.
\end{equation}

The organization of this paper is as follows.
Section~\ref{sec-classical}
introduces the features that make
the $1/\rho^{2}=1/(x^{2}+y^{2})$ potential special classically. Finding analytical solutions of the equations of
motion is
easy, and there are relatively few bound, stable orbits, but those that do exist close. The simplicity of the classical
problem
leads us to suspect that the corresponding quantum mechanical theory may also exhibit special properties.

In section~\ref{sec-separations}, we demonstrate that the Hamiltonian with a
$1/\rho^{2}$ potential belongs to the elite family that are
separable in multiple coordinate systems---cylindrical and parabolic, in this case. The scattering state wave
functions of this potential in two space dimensions have already been studied~\cite{ref-bawin}, but it is only with the
inclusion of the third dimension that the dual separations become possible. So new symmetry phenomena are expected to be
found in the three-dimensional axisymmetric potential, although it is not really clear in advance what special behavior
could be
expected. If the potential possessed bound states, we would obviously expect them to have additional degeneracies.
Unfortunately,
however, the attractive $1/\rho^{2}$ potential does not support a stable spectrum of bound negative-energy states; the
attractive
singularity at $\rho=0$ is too strong. For the scattering states of a repulsive potential, we might hope, based on the
behavior of
other multiply separable Hamiltonian systems, to uncover some interesting behavior.
The special features of the $1/\rho^{2}$ potential that we have identified are discussed in~\ref{sec-special},
and our conclusions are summarized in~section~\ref{sec-concl}.

\section{Classical Features of the $1/\rho^{2}$ Potential}
\label{sec-classical}

Central potential problems in three (or more) dimensions may be reduced to two-di\-men\-sio\-nal problems using the conservation of
angular momentum. The differential equation for the orbital curve $r(\phi)=[u(\phi)]^{-1}$ then has the simple,
well-known form
\begin{equation}
\label{eq-uofphi}
\frac{d^{2}u}{d\phi^{2}}=-u-\frac{M}{L^{2}u^{2}}F\left(\frac{1}{u}\right),
\end{equation}
where $F(r)=-dV/dr$ is the radial force. The $u$ term on the right-hand side of (\ref{eq-uofphi}) corresponds to the 
centrifugal term in the effective potential governing the radial motion. For potentials $V\propto r^{n}$, the orbital
shape can be expressed in terms of elementary trigonometric functions for $n=2$, $0$, $-1$, or $-2$~\cite{ref-bertrand}.
These are, respectively,
the harmonic oscillator, the free particle, the Coulomb potential, and the inverse square potential of interest here. For
certain other integral and rational values of the exponent $n$, the solutions may be expressed in terms of elliptic,
hypergeometric, or other progressively more general functions.

Because it is straightforward to reduce the central force problem from three to two dimensions,
there is very little practical difference at the classical level between computations with a three-dimensional central
potential $V(r)$ and the two-dimensional analogue $V(\rho)$. (The same cannot be said in quantum mechanics though; for
example, a two-dimensional
attractive potential always possesses at least one bound state, but in three dimensions the bound state need not be present.)
The classical radial equation of motion in two-dimensional space, with a potential $V(\rho)=\kappa/\rho^{2}$, using the
effective potential (depending on the angular momentum $L=L_{z}$), is
\begin{eqnarray}
\label{eq-rho-ddot}
M\frac{d^{2}\rho}{dt^{2}} & = & -\frac{d}{d\rho}\left(\frac{\kappa}{\rho^{2}}+\frac{L^{2}}{2M\rho^{2}}\right) \\
& = & \frac{2\kappa+L^{2}/M}{\rho^{3}}.
\end{eqnarray}
This is clearly just as solvable with $\kappa\neq 0$ as it is for the free particle ($\kappa=0$) case. The equation
for the orbital curve is correspondingly
\begin{equation}
\label{eq-rho-orbit}
\frac{d^{2}u}{d\phi^{2}}=-\left(1+\frac{2\kappa M}{L^{2}}\right)u.
\end{equation}
The full solutions in three dimensions are simply the two-dimensional $(\rho,\phi)$ motion superimposed upon uniform
motion in the $z$-direction, $z(t)=z(0)+\dot{z}(0)t$. The classical approach we are using corresponds to the solution
of the Schr\"{o}dinger problem in cylindrical coordinates.

The nature of the solutions for $\rho(\phi)=[u(\phi)]^{-1}$ depends on the sign of $1+2\kappa M/L^{2}$, and thus on the
sign of $\kappa$. If $\kappa>0$, the coefficient in parentheses on the right-hand side of (\ref{eq-rho-orbit}) is
automatically positive. The only possible trajectories in this repulsive potential are scattering orbits
\begin{equation}
\rho=A\sec\left(\sqrt{1+2\kappa M/L^{2}}\,\phi+\delta\right).
\end{equation}
In a coordinate system with the phase angle $\delta=0$, the radial coordinate diverges at
$\phi=\pm\pi/2\sqrt{1+2\kappa M/L^{2}}$, corresponding to a classical scattering angle
\begin{equation}
\label{eq-class-scat}
\varphi_{{\rm scat}}=\pi\left|1-\frac{1}{\sqrt{1+2\kappa M/L^{2}}}\right|,
\end{equation}
which depends on the angular momentum, but not separately on the energy---a consequence of the scale invariance
of the problem. This expression can also be cast in terms of
the impact parameter $b$ in the $xy$-plane, via $L=\sqrt{2M{\cal E}'}b$,
where ${\cal E}'>0$ is the energy of the in-plane motion (so that the total energy is ${\cal E}={\cal E}'+
\frac{1}{2}M\dot{z}^{2}$).
For attractive potentials with $\kappa<0$, there are similar scattering orbits when the energy is positive (which
means $L^{2}>2|\kappa|M$). The scattering angle is again given by (\ref{eq-class-scat}); the absolute value present in that
formula, which was superfluous for the repulsive potential, is needed in the attractive case to give a nonnegative
$\varphi_{{\rm scat}}$.

The angle $\varphi_{{\rm scat}}$ represents the scattering angle in the $xy$-plane. When the uniform motion in the
third dimension is included, it is also possible to describe the total scattering angle $\vartheta_{{\rm scat}}$.
Since the potential in three dimensions is not spherically symmetric, the scattering behavior does not depend solely on
an impact parameter (or equivalently, for fixed energy, on an angular momentum). Instead, we shall describe the incoming
trajectory of a particle by a direction $\hat{\Theta}$, together with the angular momentum component $L=L_{z}$.
Choosing an appropriate orientation for the $x$ and $y$ coordinates, $\hat{\Theta}$ is
\begin{equation}
\hat{\Theta}=\frac{1}{\sqrt{2{\cal E}'/M+\dot{z}^{2}}}\left(
\sqrt{2{\cal E}'/M}\,\hat{x}+\dot{z}\,\hat{z}\right).
\end{equation}
The $z$-axis around which the potential $V(\rho)$ is symmetric and the incoming trajectory (along which the particle would
travel if it were not deflected) are generally skew lines. Their distance of closest approach to one-another is given by the
in-plane impact parameter $b=L/\sqrt{2M{\cal E}'}$.
After the scattering, the in-plane component of the velocity has been rotated through an angle $\pm\varphi_{{\rm scat}}$,
making the outgoing direction vector
\begin{equation}
\hat{\Theta}'=\frac{1}{\sqrt{2{\cal E}'/M+\dot{z}^{2}}}\left(
\sqrt{2{\cal E}'/M}\cos\varphi_{{\rm scat}}\,\hat{x}\pm\sqrt{2{\cal E}'/M}\sin\varphi_{{\rm scat}}\,\hat{y}
+\dot{z}\,\hat{z}\right).
\end{equation}
Therefore, the three-dimensional scattering angle $\vartheta_{{\rm scat}}$ is
\begin{equation}
\vartheta_{{\rm scat}}=\cos^{-1}\left(\hat{\Theta}\cdot\hat{\Theta}'\right)=
\cos^{-1}\left(\cos\varphi_{{\rm scat}}+\frac{2\dot{z}^{2}}{2{\cal E}'/M+\dot{z}^{2}}\sin^{2}
\frac{\varphi_{{\rm scat}}}{2}\right),
\end{equation}
where $\varphi_{{\rm scat}}$ is still a function of $L$ or $b$, according to (\ref{eq-class-scat}).
Naturally, when the motion is planar ($\dot{z}=0$), this gives $\vartheta_{{\rm scat}}=\varphi_{{\rm scat}}$.
Conversely, when the velocity in the $z$-direction (which does not change) predominates,
$\vartheta_{{\rm scat}}\rightarrow0$.

Apart from the sign difference inside the absolute value in (\ref{eq-class-scat}),
there is another important difference between the attractive and repulsive regimes. When $\kappa>0$, the scattering angle is
limited to the range $0\leq\varphi_{{\rm scat}}<\pi$; the trajectory never crosses itself. In contrast, when $\kappa<0$,
the angle $\varphi_{{\rm scat}}$ may be arbitrarily large. When the potential is attractive, the particle may orbit around
the center any
number of times before it escapes again to infinity. The resulting
two-dimensional trajectories cross over themselves repeatedly. This is quite
different than the behavior seen in classical Rutherford scattering, in which the trajectories for attractive and repulsive
potentials are represented by the two disjoint branches of the same hyperbola. However, this behavior, with the number of
times
the orbital curve intersects itself increasing as the total energy approaches zero, is by no means unique to the
attractive $1/\rho^{2}$ potential, but is in fact fairly generic.

The $\kappa<0$
scattering orbits, with more and more revolutions around the origin as the energy decreases, are approaching the limit of
perfectly circular orbits, which occur when the energy vanishes at $L^{2}=-2\kappa M$. Any attractive potential will have
classical
circular orbits. However, for the potential we are interested in, it turns out that these circular orbits are, in a
certain meaningful
sense, the only bound orbits. If the total energy is negative, then the quantity in parentheses in (\ref{eq-rho-orbit}) is
negative, and the orbital solution $u(\phi)$ becomes a linear combination of equiangular spirals, so that
\begin{equation}
\label{eq-rho-sech}
\rho(\phi)=A\sech\left(\sqrt{2|\kappa|M/L^{2}-1}\,\phi+\delta\right),
\end{equation}
or
\begin{equation}
\label{eq-rho-csch}
\rho(\phi)=A\csch\left(\sqrt{2|\kappa|M/L^{2}-1}\,\phi+\delta\right),
\end{equation}
depending on whether the both endpoints lie at $\rho=0$ or one at $\rho=0$ and the other at $\rho=\infty$. Note that
all the bound orbits in two dimensions do
therefore (in a certain sense) close, because the circular orbits are the only persistent bound orbits.

Alternatively, taking the negative-energy solutions of (\ref{eq-rho-orbit}) as linear combinations
\begin{equation}
u(\phi)=B\cosh\left(\sqrt{2|\kappa|M/L^{2}-1}\,\phi\right)+C\sinh\left(\sqrt{2|\kappa|M/L^{2}-1}\,\phi\right),
\end{equation}
$\rho(\phi)$ has the form (\ref{eq-rho-sech}), with both endpoints at the center, if $|B|>|C|$; it has the form
(\ref{eq-rho-csch}) if $|B|<|C|$. The intermediate cases, with $B=\pm C$, yields pure inward or outward equiangular spirals. 
The spirals are also limiting forms of the other expressions, with $\delta\rightarrow\infty$ while
$Ae^{-\delta}$ is held finite.

Most generally, for states with $\kappa+L^{2}/2M<0$,
energy ${\cal E}'<0$, and initial radial velocity inward [$\dot{\rho}(0)\leq0$ at time $t=0$], the equation of motion
(\ref{eq-rho-ddot}) has the implicit solution
\begin{equation}
t=\frac{\rho}{2|{\cal E}'|}\sqrt{\frac{|2\kappa M+L^{2}|}{\rho^{2}}-2M|{\cal E}'|}+
\frac{M\rho(0)\dot{\rho}(0)}{2|{\cal E}'|}.
\end{equation}
In this regime, the time $t_{f}$ required for the particle to reach $\rho=0$ is
\begin{equation}
t_{f}=\frac{\sqrt{|2\kappa M+L^{2}|}}{2|{\cal E}'|}-\frac{M\rho(0)|\dot{\rho}(0)|}{2|{\cal E}'|}
\end{equation}
This is behavior is known as ``falling to the center.''
Note that making $\dot{\rho}(0)$ more negative while keeping $\rho(0)$ fixed decreases $t_{f}$, which is clearly correct;
the effects of varying $\rho(0)$ independently are less intuitively obvious.

\section{Two Separations of the Schr\"{o}dinger Equation}
\label{sec-separations}

We now turn our attention to the quantum theory.
The classical theory with an attractive potential was dominated by the falling to the center.
The quantum mechanical wave functions in the presence of the $\kappa<0$ potential exhibit their own manifestation of this
phenomenon. If $\psi(\vec{\rho}\,)$ is an eigenfunction of the two-dimensional
time-independent Schr\"{o}dinger equation (with energy eigenvalue ${\cal E}'_{0}$), then for
any real number $\alpha$, $\psi(\alpha\vec{\rho}\,)$ is also an eigenfunction, with energy $\alpha^{2}{\cal E}'_{0}$. Thus,
if there is a
normalizable eigenstate with energy ${\cal E}'_{0}<0$, then there must be eigenstates with arbitrarily negative energies. By
compressing
the wave function closer to the attractive singularity at $\rho=0$, the energy may be made as negative as we wish, meaning
that there cannot be a stable Hilbert space of quantum states; the energy is not bounded below.

In contrast, when $\kappa>0$, the energy eigenvalues are always
positive. We may still dilate the wave function to decrease its energy, but the energies remain bounded from below by zero;
and it is, of course, no surprise that all positive energies are allowed in this scattering system.

It is possible, via one of several renormalization procedures, to let the strength of the attractive $1/r^{2}$ potential
go to zero,
in such a way that there is a reasonable physical spectrum (with exactly one bound
state)~\cite{ref-camblong,ref-coon,ref-bouaziz}. However, the resulting
Hamiltonian is not self-adjoint, in spite of it having a naively Hermitian appearance~\cite{ref-case};
moreover, the scale invariance of the
solutions is broken by an anomaly. Most analyses of the regularized Hamiltonian have focused on the three-dimensional
$1/r^{2}$ potential, although the general character of the solutions appears to be independent of the
dimensionality~\cite{ref-camblong,ref-essin,ref-avila,ref-burgess}.
Our results in this paper might be extended to this renormalized regime; in fact,
it would be very interesting to see how the renormalization would interact with the separation of the Schr\"{o}dinger
equation in parabolic coordinates. However, this regime lies beyond the scope of the present work.

We shall now present separation of variables solutions to the Schr\"{o}dinger equation for the $1/\rho^{2}$ potential
in both cylindrical and parabolic coordinates. The solution in cylindrical coordinates is quite conventional. It represents
a straightforward application of well-known techniques, and the solutions involve the same families of special functions
that are commonplace in two-dimensional problems. However, the solution in parabolic coordinates is completely new. It may
be surprising that the Schr\"{o}dinger equation involved is separable in parabolic coordinates at all, particularly since
the use of parabolic coordinates breaks the translational symmetry---which plays such an important role in
cylindrical coordinates---of the problem. Moreover, the separation solution in parabolic coordinates will also provide
insight into an alternative method for solving the classical problem that was considered in
section~\ref{sec-classical}.

\subsection{Cylindrical Coordinates}

In order to have a quantum theory with well-defined wave functions,
without additional regularization of the potential in the vicinity of $\rho=0$,
the potential we shall consider in the remainder of our analysis is
\begin{equation}
V(\vec{r}\,)=\frac{\hbar^{2}K}{2M}\frac{1}{\rho^{2}},
\end{equation}
with repulsive $K>0$. (The potential strength $\kappa$ has been rescaled to avoid unnecessary factors of $\hbar$ and $M$.)
With this potential, the Schr\"{o}dinger equation in cylindrical coordinates is
\begin{equation}
\left[-\frac{1}{\rho}\frac{\partial}{\partial\rho}\left(\rho\frac{\partial}{\partial\rho}\right)-
\frac{1}{\rho^{2}}\frac{\partial^{2}}{\partial\phi^{2}}
-\frac{\partial^{2}}{\partial z^{2}}+\frac{K}{\rho^{2}}\right]\psi=E\psi,
\end{equation}
where $E=2M{\cal E}/\hbar^{2}$, with ${\cal E}$ being the total energy. With a separable ansatz,
\begin{equation}
\psi=P(\rho)e^{ikz}e^{im\phi},
\end{equation}
this reduces to a single-variable Schr\"{o}dinger equation for $P(\rho)$,
\begin{equation}
\left[-\frac{1}{\rho}\frac{d}{d\rho}\left(\rho\frac{d}{d\rho}\right)+\frac{m^{2}+K}{\rho^{2}}\right]P=\left(E-k^{2}\right)P.
\end{equation}
This is just the usual Bessel's equation that arises for a free particle in two dimensions, except with the indices of the
Bessel function solutions changed to $\sqrt{m^{2}+K}$. (Note that $m^{2}+K$ just corresponds to the classical quantity
$L_{z}^{2}+2M\kappa$, measured in units of $\hbar^{2}$.)
The general solution is thus
\begin{equation}
P(\rho)=AJ_{\sqrt{m^{2}+K}}\left(\sqrt{E-k^{2}}\,\rho\right)+BJ_{-\sqrt{m^{2}+K}}\left(\sqrt{E-k^{2}}\,\rho\right).
\end{equation}
If the Bessel function index $\sqrt{m^{2}+K}$ happens to be an integer, the usual replacement of the
linearly dependent $J_{-\sqrt{m^{2}+K}}$ by the Neumann function $N_{\sqrt{m^{2}+K}}$ is required.
However, only the Bessel function with positive index is regular at $\rho=0$ (and thus permitted).

The scattering theory of these solutions is straightforward. The scattering by a three-dimensional $1/r^{2}$ potential
is worked out in~\cite{ref-kayser}. One surprising result is that the classical limit only exists for strong potentials,
as the classical scattering cross section is linear (never quadratic) in the strength of the potential.
The two-dimensional $1/\rho^{2}$ version is completely analogous, merely using the
formalism for partial wave scattering in two dimensions~\cite{ref-lapidus}.
The partial wave expansion for an incoming plane wave is
\begin{equation}
e^{iqx}=J_{0}(q\rho)+2\sum_{m=1}^{\infty}i^{m}\cos(m\phi)J_{m}(q\rho),
\end{equation}
in terms of the free wave radial functions.
The Bessel functions have the limiting behavior $J_{\nu}(s)=\sqrt{\frac{2}{\pi s}}\cos\left(s-\frac{\nu\pi}{2}
-\frac{\pi}{4}\right)$, and the scattering state wave function may be written
\begin{eqnarray}
\psi & = & e^{iq\rho}+\psi_{{\rm scat}} \\
& \rightarrow & e^{iq\rho}+\sqrt{\frac{2}{\pi q\rho}}\left[\cos\left(q\rho-\frac{\pi}{4}+\delta_{0}
\right)+2\sum_{m=1}^{\infty}i^{m}\cos(m\phi)\cos\left(q\rho-\frac{m\pi}{2}-\frac{\pi}{4}+\delta_{m}\right)
\right], \nonumber
\end{eqnarray}
where $q=\sqrt{E-k^{2}}$.
The non-free wave functions with $J_{\sqrt{m^{2}+K}}(q\rho)$ are simply phase shifted by
\begin{equation}
\delta_{m}=-\frac{\pi}{2}\left(\sqrt{m^{2}+K}-|m|\right).
\end{equation}
The fact that $\delta_{m}$ is independent of the energy for each partial wave is another
consequence of the scale invariance. Moreover, as noted, the classical limit corresponds to $K\gg|m|$.

\subsection{Parabolic Coordinates}

Unlike the scattering solution in cylindrical coordinates, the solution of the $1/\rho^{2}$ potential in
parabolic coordinates is not a standard problem. A $1/\rho^{2}$ potential has, however,
previously been considered algebraically, as a perturbation
added to the Coulomb Hamiltonian (which, as noted above, is also parabolic separable)~\cite{ref-helfrich,ref-kais}.

In parabolic coordinates, the Laplacian is
\begin{equation}
\vec{\nabla}^{2}=\frac{4}{\eta+\xi}\frac{\partial}{\partial\eta}\left(\eta\frac{\partial}{\partial\eta}\right)+
\frac{4}{\eta+\xi}\frac{\partial}{\partial\xi}\left(\xi\frac{\partial}{\partial\xi}\right)+
\frac{1}{\eta\xi}\frac{\partial^{2}}{\partial\phi^{2}}.
\end{equation}
Whether an eigenfunction is separable in cylindrical coordinates, with $\psi=P(\rho)\Phi(\phi)Z(z)$
or in parabolic coordinates $\psi=H(\eta)\Xi(\xi)\Phi(\phi)$, we may take it to be an eigenfunction of
$L_{z}$, $\Phi(\phi)=e^{im\phi}$. Noting that $\eta\xi=\rho^{2}$, taking this azimuthal dependence reduces the
Laplacian plus potential in the parabolic coordinates to
\begin{eqnarray}
-\vec{\nabla}^{2}+\frac{K}{\rho^{2}} & = & -\frac{4}{\eta+\xi}\frac{\partial}{\partial\eta}
\left(\eta\frac{\partial}{\partial\eta}\right)-
\frac{4}{\eta+\xi}\frac{\partial}{\partial\xi}\left(\xi\frac{\partial}{\partial\xi}\right)+\frac{m^{2}+K}{\eta\xi} \\
& = &  \frac{1}{\eta+\xi}\left[-4\frac{\partial}{\partial\eta}\left(\eta\frac{\partial}{\partial\eta}\right)-
4\frac{\partial}{\partial\xi}\left(\xi\frac{\partial}{\partial\xi}\right)+(m^{2}+K)\left(\frac{1}{\eta}+\frac{1}{\xi}\right)
\right].
\end{eqnarray}
Once again, and not coincidentally, the inclusion of the potential corresponds to the change $m^{2}\rightarrow m^{2}+K$.

With $\Phi(\phi)$ factored out, the remaining Schr\"{o}dinger equation can be written in the separation form
\begin{equation}
\label{eq-HXi-sep}
\left[\frac{4}{H}\frac{\partial}{\partial\eta}\left(\eta\frac{\partial H}{\partial\eta}\right)+E\eta
-\left(m^{2}+K\right)\frac{1}{\eta}\right]+
\left[\frac{4}{\Xi}\frac{\partial}{\partial\xi}\left(\xi\frac{\partial\Xi}{\partial\xi}\right)+E\xi
-\left(m^{2}+K\right)\frac{1}{\xi}\right]=0.
\end{equation}
Letting the first bracketed term in (\ref{eq-HXi-sep}) be equal to a constant $C$, the ordinary differential equation
for $H$ is
\begin{equation}
\label{eq-H-ODE}
4\eta^{2}\frac{d^{2}H}{d\eta^{2}}+4\eta\frac{dH}{d\eta}+\eta^{2}EH-\eta CH-\left(m^{2}+K\right)H=0,
\end{equation}
and with $C\rightarrow -C$ in the equation for $\Xi$.

Since the wave function is complex, it may not be automatically clear
whether $C$ should be real or complex. Note that a purely imaginary
$C$ gives the real and imaginary parts of the solutions to the 
ordinary differential equations definite behavior under inversions of the
variables, $\eta\rightarrow -\eta$ or $\xi\rightarrow -\xi$. However, this
behavior is not actually physically mandated by the theory, because the
physical space is limited to the parameter region where both $\eta$ and
$\xi$ are nonnegative. It will, however, necessarily be the case that only a one-parameter family of $C$ values will
correspond to physically meaningful states. Any separable energy eigenfunction in three dimensions is determined
(up to phase and normalization) by the values of the three real quantum numbers. In this system, we have the
physical observables represented by $m$ and $E$, so the choice of $C$ must provide exactly one additional real degree
of freedom. Since with a real-valued $C$, the separate differential equations for $H$ and $\Xi$ can be cast as
eigenvalue equations for Hermitian operators, a real $C$ is a sufficient condition for having equations that yield
bases of wave functions with asymptotic forms
that are continuum normalizable. Thus, a real $C$ gives the correct one-parameter family of solutions.

The linearly independent solutions of (\ref{eq-H-ODE}) are expressible in terms of the confluent hypergeometric
fuctions $_{1}F_{1}(a;b;s)$,
\begin{equation}
h_{\pm}(\eta)=\eta^{\pm\frac{1}{2}\sqrt{m^{2}+K}}e^{-\frac{i}{2}\sqrt{E}\eta}\,_{1}F_{1}\left(\frac{1}{2}\pm\frac{1}{2}
\sqrt{m^{2}+K}-\frac{iC}{4\sqrt{E}};1\pm\sqrt{m^{2}+K};i\sqrt{E}\eta\right).
\end{equation}
In order to have regularity at the origin (where $\eta=\xi=0$), we must have the solution $H=h_{+}(\eta)$, and for $\Xi$,
\begin{equation}
\Xi=h'_{+}(\xi)=\xi^{\frac{1}{2}\sqrt{m^{2}+K}}e^{-\frac{i}{2}\sqrt{E}\xi}\,_{1}F_{1}\left(\frac{1}{2}+\frac{1}{2}
\sqrt{m^{2}+K}+\frac{iC}{4\sqrt{E}};1+\sqrt{m^{2}+K};i\sqrt{E}\xi\right).
\end{equation}
Unfortunately, the overlap integrals giving the weights needed to write the wave functions
$H(\eta)\Xi(\xi)$ as superpositions
of the $P(\rho)Z(z)$ in cylindrical coordinates are intractable in the general case.

The asymptotic behavior of $_{1}F_{1}(a;b;s)$ for $|s|\rightarrow\infty$ and $-\frac{3\pi}{2}<\arg s<\frac{\pi}{2}$ is
\begin{equation}
\label{eq-1F1-asym}
_{1}F_{1}(a;b;s)\sim\Gamma(b)\left[\frac{e^{s}s^{a-b}}{\Gamma(a)}+\frac{e^{i\pi a}s^{-a}}{\Gamma(b-a)}\right].
\end{equation}
For the solutions $h_{+}$ and $h'_{+}$, the relevant values of $a$ and $b-a(=a^{*})$ always have real parts
$\frac{1}{2}+\frac{1}{2}\sqrt{m^{2}+K}$, which means that the first and second terms in (\ref{eq-1F1-asym})
are of the same magnitude when $\eta$ or $\xi$ is large. The large $\eta$ behavior of the
confluent hypergeometric function appearing in $h_{+}(\eta)$ is accordingly [using the phase convention
that $s=e^{\frac{i\pi}{2}}\sqrt{E}\eta$, corresponding to that in (\ref{eq-1F1-asym})],
\begin{eqnarray}
_{1}F_{1} & \sim & \Gamma\left(1+\sqrt{m^{2}+K}\right)\left(i\sqrt{E}\eta\right)^{-\left(\frac{1}{2}+
\frac{1}{2}\sqrt{m^{2}+K}\right)}
\left[\frac{e^{i\sqrt{E}\eta}\left(e^{\frac{i\pi}{2}}\sqrt{E}\eta\right)^{-\left(\frac{iC}{4\sqrt{E}}\right)}}
{\Gamma\left(\frac{1}{2}+\frac{1}{2}\sqrt{m^{2}+K}-\frac{iC}{4\sqrt{E}}\right)}\right. \\
& & +\left.\frac{e^{i\pi\left(\frac{1}{2}+\frac{1}{2}\sqrt{m^{2}+K}-\frac{iC}{4\sqrt{E}}\right)}
\left(e^{\frac{i\pi}{2}}\sqrt{E}\eta\right)^{\left(\frac{iC}{4\sqrt{E}}\right)}}
{\Gamma\left(\frac{1}{2}+\frac{1}{2}\sqrt{m^{2}+K}+\frac{iC}{4\sqrt{E}}\right)}\right] \nonumber\\
\label{eq-h+int}
& = & \frac{\Gamma\left(1+\sqrt{m^{2}+K}\right)
e^{\frac{\pi C}{8\sqrt{E}}}
\left(i\sqrt{E}\eta\right)^{-\left(\frac{1}{2}+\frac{1}{2}\sqrt{m^{2}+K}\right)}}
{\left|\Gamma\left(\frac{1}{2}+\frac{1}{2}\sqrt{m^{2}+K}-\frac{iC}{4\sqrt{E}}\right)\right|}
\left[e^{i\left(\sqrt{E}\eta-\arg\Gamma(a)+\frac{C}{4\sqrt{E}}\log\sqrt{E}\eta\right)}\right. \\
& & +\left.e^{i\left(\frac{\pi}{2}+\frac{\pi}{2}\sqrt{m^{2}+K}+\arg\Gamma(a)-\frac{C}{4\sqrt{E}}\log\sqrt{E}\eta\right)}
\right] \nonumber\\
\label{eq-h+final}
& = & \frac{2\Gamma\left(1+\sqrt{m^{2}+K}\right)e^{\frac{\pi C}{8\sqrt{E}}}
\left(\sqrt{E}\eta\right)^{-\left(\frac{1}{2}+\frac{1}{2}\sqrt{m^{2}+K}\right)}}
{\left|\Gamma\left(\frac{1}{2}+\frac{1}{2}\sqrt{m^{2}+K}-\frac{iC}{4\sqrt{E}}\right)\right|}e^{\frac{i}{2}\sqrt{E}\eta}
\cos\left[\frac{1}{2}\sqrt{E}\eta\right. \\
& & +\left.\frac{C}{8\sqrt{E}}\log\left(\sqrt{E}\eta\right)-
\frac{1}{2}\arg\Gamma\left(\frac{1}{2}+\frac{1}{2}\sqrt{m^{2}+K}-\frac{iC}{4\sqrt{E}}\right)-\frac{\pi}{4}\sqrt{m^{2}+K}
-\frac{\pi}{4}\right]. \nonumber
\end{eqnarray}
In the intermediate formula (\ref{eq-h+int}), $\arg\Gamma(a)$ has been used to abbreviate the complex argument of
$\Gamma\left(\frac{1}{2}+\frac{1}{2}\sqrt{m^{2}+K}-\frac{iC}{4\sqrt{E}}\right)$.

It follows that the asymptotic behavior of the wave function ($r\rightarrow\infty$,
but away from from the $z$-axis, where $\eta$ or $\xi$ will vanish) is
\begin{eqnarray}
H(\eta)\Xi(\xi) & \propto & \frac{1}{\sqrt{\eta\xi}}\cos\left[\frac{1}{2}\sqrt{E}\eta+\frac{C}{8\sqrt{E}}
\log\left(\sqrt{E}\eta\right)-\frac{1}{2}\arg\Gamma(a)-\frac{\pi}{4}\sqrt{m^{2}+K}-\frac{\pi}{4}\right] \nonumber\\
& & \times\cos\left[\frac{1}{2}\sqrt{E}\xi+\frac{C}{8\sqrt{E}}
\log\left(\sqrt{E}\xi\right)-\frac{1}{2}\arg\Gamma(a^{*})-\frac{\pi}{4}\sqrt{m^{2}+K}-\frac{\pi}{4}\right] \\
\label{eq-HX-asym}
& = & \frac{1}{2\rho}\left\{\cos\left[\sqrt{E}r+\frac{C}{8\sqrt{E}}\log\left(\cot^{2}\frac{\theta}{2}\right)
-\frac{\pi}{2}\sqrt{m^{2}+K}-\frac{\pi}{2}\right]\right. \\
& & \left.+\cos\left[\sqrt{E}z+\frac{C}{8\sqrt{E}}\log\left(E\rho^{2}\right)-\arg\Gamma(a)\right]\right\}, \nonumber
\end{eqnarray}
using $\eta\xi=\rho^{2}$ and $\frac{\eta}{\xi}=\cot^{2}\frac{\theta}{2}$.

The limiting form (\ref{eq-HX-asym}) away from the $z$-axis is clearly normalizable as a continuum state. On the other
hand, in the vicinity of the $z$-axis, either $h_{+}(\eta)$ or $h_{+}'(\xi)$ is close to 1, while the other
function---and the wave function
$\psi$ as a whole---scale as $\sim1/\sqrt{|z|}$, which is again normalizable behavior. This confirms
that our earlier choice of a real separation constant $C$ was the correct one for the physical wave function
solutions.


\section{Special Features}
\label{sec-special}

Remarkably, the separated equation (\ref{eq-H-ODE}) for $H$ can actually be cast in
nearly the same form as the radial
Schr\"{o}dinger equation for a Coulomb potential. Letting $U_{1}(\eta)=\sqrt{\eta}H(\eta)$, (\ref{eq-H-ODE})
becomes
\begin{equation}
\label{eq-coul-like}
-\frac{d^{2}U_{1}}{d\eta^{2}}+\frac{C}{4\eta}U_{1}+\frac{m^{2}+K-1}{4\eta^{2}}U_{1}=\frac{E}{4}U_{1}.
\end{equation}
The ordinary differential equation of $U_{2}(\xi)=\sqrt{\xi}\Xi(\xi)$ is identical, except for the switch
$C\rightarrow-C$, equivalent to interchanging an attractive Coulomb potential with a repulsive one.
Moreover, the normalization condition for the wave function,
\begin{equation}
\frac{1}{4}\int_{0}^{\infty}d\xi\int_{0}^{\infty} d\eta\,(\eta+\xi)\left|H(\eta)\Xi(\xi)\right|^{2}=
\frac{1}{4}\int_{0}^{\infty}d\xi\int_{0}^{\infty} d\eta\,\left(\frac{1}{\eta}+\frac{1}{\xi}\right)
\left|U_{1}(\eta)U_{2}(\xi)\right|^{2}=\frac{1}{2\pi},
\end{equation}
sets the same kinds of constraints on how quickly
the functions $U_{1}$ and $U_{2}$ must decay at spatial infinity as in the Coulomb problem.
The equivalence
also immediately explains the presence of the $\log(\sqrt{E}\eta)$ and $\arg\Gamma(a)$ terms in the argument of
the cosine in (\ref{eq-h+final}), since these same
kinds of terms appear in the phases of Coulomb waves.

The transformation of the separated parts of the Schr\"{o}dinger equation into Coulomb-like forms opens up
a number of tools that can be used to further analyze the wave function solutions. However, those tools may
play different roles in the analysis of the $1/\rho^{2}$ potential than in the study of the $1/r$ potential.
For example, there are $(0+1)$-dimensional supersymmetry transformations that carry solutions of the radial
Schr\"{o}dinger equation in the Coulomb problem to other radial solutions with the same energies but different
values of the angular momentum (changing $l \leftrightarrow l+1$)~\cite{ref-cooper1,ref-rajagopal}.
Applied to (\ref{eq-coul-like}), these transformations would still leave the energy unaffected (and also the
separation constant $C$ unchanged), but the strength of the potential would be modified through a change to the
quantity $m^{2}+K$, which combines the $z$-component of angular momentum with the strength of the repulsive potential.
This is analogous to the situation with the one-dimensional $\sech^{2}ax$ potential, where the supersymmetry
transformations connect potentials with the same functional form, but of different depths.

Another interesting feature of the solution in parabolic coordinates stems from the fact that the
choice of coordinate system breaks the translation invariance along the $z$-direction. This symmetry is
manifestly present in the equations of motion in a cylindrical coordinate system, where $z$ is a cyclic coordinate.
Since $z$ never enters the dynamics explicitly, there is nothing special about the location of $z=0$. The
translation symmetry is obscured somewhat in the parabolic coordinates, but it must still exist. If
$\psi_{1}=H(\eta)\Xi(\xi)e^{im\phi}$ is a solution of the Schr\"{o}dinger equation, then
\begin{equation}
\psi_{2}=H\left[\sqrt{\rho^{2}+(z-a)^{2}}+(z-a)\right]\Xi\left[\sqrt{\rho^{2}+(z-a)^{2}}-(z-a)\right]e^{im\phi}
\end{equation}
must also be a solution, since it is simply a translate of $\psi_{1}$ along the $z$-direction. The degeneracy
of these states is analogous to the energy degeneracy of the Landau levels for a charged particle moving in the
plane perpendicular to a constant magnetic field. The magnitude of the degeneracy is proportional to the area
of the plane, since the origin of the coordinates may be located anywhere in the plane.

The presence of $C$ in the eigenfunction equation (\ref{eq-coul-like}) also appears to break the scale invariance
of the problem, since the Hermitian operator on the left-hand side contains $C$, which has units of (length)$^{-1}$.
However, since $C$ is merely a separation constant, which can take any real value, a rescaling
$\vec{\rho}\rightarrow\alpha\vec{\rho}$ (and thus $\eta\rightarrow\alpha\eta$, $\xi\rightarrow\alpha\xi$)
may be accompanied by $C\rightarrow\alpha^{-1}C$. Since the same $C$ appears in the equations for $H$ and $\Xi$,
this restores the physical scaling invariance.

Finally, the separation of the quantum-mechanical problem in parabolic coordinates can give some insight about the
classical behavior in that coordinate system. Because the $\eta$ and $\xi$ portions of the Schr\"{o}dinger equation
are the same as those for two Coulomb problems, one attractive and the other equally repulsive, we can apply the normal
methods for solving the Kepler problem to the classical time evolution of a particle's $(\eta,\xi,\phi)$ coordinates.
Recalling the that the classical limit applies when $K$ is large, we may neglect the $-1$ in the $m^{2}+K-1$
appearing in (\ref{eq-coul-like}). Then, restoring the factors of $\hbar^{2}/2M$, (\ref{eq-coul-like})
corresponds to a classical limit of
\begin{equation}
\label{eq-eta-class}
\frac{1}{2}M\dot{\eta}^{2}+\frac{{\cal C}}{4\eta}+\frac{L^{2}+2M\kappa}{4\eta^{2}}=\frac{{\cal E}}{4},
\end{equation}
where $C=2M{\cal C}/\hbar^{2}$.
There is no orbital equation that is directly analogous to the one for $r(\phi)$ in the normal Kepler problem, since the
relationship between the angular velocity $\dot{\phi}$ and $L=L_{z}=M\rho^{2}\dot{\phi}=M\eta\xi\dot{\phi}$ is determined
by both $\eta$ and $\xi$ together. However, (\ref{eq-eta-class}) may be solved implicitly for the time $t$ as a function
of the coordinate $\eta$,
\begin{eqnarray}
t & = & \sqrt{\frac{M}{2}}\int_{\eta_{{\rm min}}}^{\eta}\frac{d\eta'}{\sqrt{\frac{{\cal E}}{4}-\frac{{\cal C}}{4\eta'}
-\frac{L^{2}+2M\kappa}{8(\eta')^{2}}}} \\
\label{eq-time-eta}
& = & \frac{\sqrt{2M{\cal E}\eta^{2}-2M{\cal C}\eta-L^{2}-2M\kappa}}{{\cal E}} \\
& & +\frac{\sqrt{M}{\cal C}}
{\sqrt{2}{\cal E}^{\frac{3}{2}}}
\log\left[2\sqrt{{\cal E}\left({\cal E}\eta^{2}-{\cal C}\eta-\frac{L^{2}}{2M}-\kappa\right)}
+2{\cal E}\eta-{\cal C}\right] \nonumber\\
& & -\frac{\sqrt{M}{\cal C}}{2\sqrt{2}{\cal E}^{\frac{3}{2}}}
\log\left[{\cal C}^{2}+4{\cal E}\left(\frac{L^{2}}{2M}+\kappa\right)\right]. \nonumber
\end{eqnarray}
The origin of the time coordinate has been chosen in this case so that $t=0$ occurs at the turning point
\begin{equation}
\eta(0)=\eta_{{\rm min}}=\frac{{\cal C}+\sqrt{{\cal C}^{2}+4{\cal E}\left(\frac{L^{2}}{2M}+\kappa\right)}}{2{\cal E}}
\end{equation}
for $\eta$. Since condition for $\eta_{{\rm min}}$ is
$\sqrt{2M{\cal E}\eta_{{\rm min}}^{2}-2M{\cal C}\eta_{{\rm min}}-L^{2}-2M\kappa}=0$, both square roots in
(\ref{eq-time-eta}) vanish at $\eta=\eta_{{\rm min}}$, and just the last term comes from the lower limit of the integration.

Simultaneously, the $\xi$ coordinate is evolving independently. The time is once again given implicitly, in this
instance by
\begin{eqnarray}
t & = & \frac{\sqrt{2M{\cal E}\xi^{2}+2M{\cal C}\xi-L^{2}-2M\kappa}}{{\cal E}}
-\frac{\sqrt{2M{\cal E}\xi(0)^{2}+2M{\cal C}\xi(0)-L^{2}-2M\kappa}}{{\cal E}} \\
& & -\frac{\sqrt{M}{\cal C}}
{\sqrt{2}{\cal E}^{\frac{3}{2}}}
\log\left\{\frac{2\sqrt{{\cal E}\left({\cal E}\xi^{2}+{\cal C}\xi-\frac{L^{2}}{2M}-\kappa\right)}
+2{\cal E}\xi+{\cal C}}{2\sqrt{{\cal E}\left[{\cal E}\xi(0)^{2}+{\cal C}\xi(0)-\frac{L^{2}}{2M}-\kappa\right]}
+2{\cal E}\xi(0)+{\cal C}}\right\}. \nonumber
\end{eqnarray}
$\xi(0)$ is the value of $\xi$ when $\eta=\eta_{{\rm min}}$. If $\eta(t)$ and $\xi(t)$ are determined, then the
remaining angular behavior can be found from
\begin{equation}
\phi(t)=\phi(0)+\frac{L}{M}\int_{0}^{t}\frac{dt'}{\eta(t')\xi(t')},
\end{equation}
completing the classical solution.

\section{Conclusions}
\label{sec-concl}

Hamiltonians that are amenable to separation of variables methods in more that one coordinate system have a number
of important properties. These include accidental degeneracies in their bound state spectra, and classical behavior that
typically involves bound orbits that always close. The nonrelativistic Hamiltonians for number of important physical
systems, such as the hydrogen atom and the charged particle in a constant magnetic field, are multiply separable in this
way. These features are also tied to the usefulness of operator methods in solving these Hamiltonians.

The $z$-independent $1/\rho^{2}$ potential in three dimensions is obviously separable in cylindrical coordinates
$(\rho,\phi,z)$, and we have shown that it is also separable in parabolic coordinates $(\eta,\xi,\phi)$. Although parabolic
coordinates are not used nearly as frequently as rectangular, spherical, and cylindrical coordinate systems, they
were already known to be useful for addressing certain aspects of the Coulomb problem. The attractive
$1/\rho^{2}$ potential is too strong to support a stable set of bound states, but the repulsive version is well behaved.
When separated in parabolic coordinates, the one-dimensional Schr\"{o}dinger equations for the component functions
$H(\eta)$ and $\Xi(\xi)$ have the same forms as the radial Schr\"{o}dinger equation in the Coulomb problem, although
the strength of the Coulomb-like term is set by the separation constant $C$, so that one equation features the effective
equivalent of an attractive potential, which the other has a repulsive potential of equal magnitude.

The asymptotic behavior of the of full wave function $\psi(\eta,\xi,\phi)$ in parabolic coordinates is thus determined by
the limiting behavior of a product of Coulomb waves, albeit ones that typically have nonintegral values for what would
normally be the angular momentum parameter $\ell$. Moreover, although we have focused on the scattering states of a
repulsive $1/\rho^{2}$ potential, it is clear that for sufficiently large values of $L_{z}=m\hbar$, the scattering
wave function in an attractive $1/\rho^{2}$ potential will have essentially the same structures. They will still be
products of Coulomb radial functions in $\eta$ and $\xi$, with unconventional values of $\ell$ and equal and opposite
effective potential strengths.

While the separation of variables in cylindrical coordinates keeps the translation symmetry along the $z$-direction
and the scale invariance of the problem manifest, both of these invariances are obscured in the parabolic coordinate
system, which picks a particular $z=0$ location about with the $\eta$ and $\xi$ coordinate surfaces are focused. The
fact that these important features are hidden in the parabolic coordinates formalism suggests that there may be yet other
interesting features of these potentials
still to be uncovered. In any case, the planar $1/\rho^{2}$ potential, as well as being
a system of real physical significance~\cite{ref-denschlag,ref-nowak,ref-strebel},
appears to be a fruitful theoretical laboratory for
understanding the structure of mechanics in parabolic coordinates and the behavior of multiply separable quantum systems.

\end{document}